\documentclass[]{aa}
\usepackage{natbib}
\bibpunct{(}{)}{;}{a}{}{,}
\usepackage{graphicx}
\usepackage{txfonts}
\begin{document}

\newcommand{\Rsun}{$R_{\odot}$}
\newcommand{\kms}{km~s$^{-1}$}
\newcommand{\Msun}{$M_\odot$}
\newcommand{\Teff}{$T_{\rm eff}$}
\newcommand{\CS}{CS~30322-023}
\newcommand{\HE}{HE~0017+0055}

\titlerunning{Properties of the CEMP-rs star HE~0017+0055}
\title{HE~0017+0055 : A probable pulsating CEMP-rs star and long-period binary
\thanks{Based on observations performed with the Mercator telescope and the Nordic Optical Telescope (NOT), operated by the Nordic Optical Telescope Scientific Association at the Roque de los Muchachos Observatory, La Palma, Spain, of the Instituto de Astrof\"{i}sica de Canarias.}}
  
\author{A. Jorissen\inst{1}
\and T. Hansen\inst{2} 
\and S. Van Eck\inst{1}
\and J. Andersen \inst{3,4} 
\and B. Nordstr\"{o}m \inst{3,4}
\and L. Siess \inst{1}
\and G. Torres \inst{5}
\and T. Masseron \inst{6}
\and H. Van Winckel \inst{7}
          }
\offprints{A. Jorissen}

\institute{Institut d'Astronomie et d'Astrophysique, Universit\'e Libre de
Bruxelles, CP 226, Boulevard du Triomphe, B-1050 Bruxelles, Belgium \\
       \email{ajorisse@ulb.ac.be,svaneck@ulb.ac.be,lsiess@ulb.ac.be} 
\and
Landessternwarte, ZAH, Heidelberg University,
              K{\"o}nigstuhl 12, Heidelberg, D-69117, Germany\\
              \email{thansen@lsw.uni-heidelberg.de}
   \and 
             Dark Cosmology Centre, The Niels Bohr Institute, University of Copenhagen,
             Juliane Maries Vej 30, DK-2100 Copenhagen, Denmark\\
             \email{ja@astro.ku.dk, birgitta@astro.ku.dk}
\and
Stellar Astrophysics Centre, Department of Physics and Astronomy, Aarhus 
University, DK-8000 Aarhus C, Denmark 
\and
Harvard-Smithsonian Center for Astrophysics\\ 
60 Garden Street, Cambridge, MA 02138, USA\\
\email{gtorres@cfa.harvard.edu}
\and
Institute of Astronomy, University of Cambridge, Madingley Road, Cambridge CB3 0HA, UK\\
\email{tpm40@ast.cam.ac.uk}
\and
Instituut voor Sterrekunde, Katholieke Universiteit Leuven \\ 
Celestijnenlaan 200d - bus 2412, BE - 3001 Heverlee, Belgium\\
\email{Hans.VanWinckel@ster.kuleuven.be}
}

\date{Received  /  Accepted }

\abstract{A large fraction of the carbon-enhanced extremely metal-poor halo giants ([Fe/H] $<~ -2.5$) are also strongly enriched in neutron-capture elements from the $s$-process (CEMP-s stars). The conventional explanation for the properties of these stars is mass transfer 
from a nearby binary companion on the Asymptotic Giant Branch (AGB). This scenario leads to a number of testable predictions in terms of the properties of the putative binary system and the resulting abundance pattern.  Some among the CEMP stars further exhibit overabundances in r-process elements on top of the s-process enrichment, and are tagged CEMP-rs stars. Although the nucleosynthesis process responsible for such a mixed abundance pattern is still debated, these stars seem to belong to binary systems as do CEMP-s stars.
}
{Our aim is to present and analyse in detail our comprehensive data set of systematic radial-velocity measurements and high-resolution spectroscopy of the CEMP  star \HE.
} 
{Our precise radial-velocity monitoring of \HE\ over 2940 days (8 yr) with the Nordic Optical Telescope  and Mercator telescopes exhibits variability with a period of 384~d and amplitude of $540\pm27$~m~s$^{-1}$, superimposed on a nearly linear long-term decline of $\sim$1~m~s$^{-1}$~day$^{-1}$. High-resolution HERMES/Mercator and Keck/HIRES spectra have been used to derive elemental abundances using 1-D LTE MARCS models.  A metallicity of [Fe/H]~$\sim -2.4$ is found, along with s-process overabundances on the order of 2~dex (with the exception of [Y/Fe]~$\sim+0.5$),  and most notably overabundances 
of r-process elements like Sm, Eu, Dy, and Er in the range 0.9 -- 2.0~dex. With [Ba/Fe]~$ > 1.9$~dex and [Eu/Fe]~=~2.3~dex, \HE\ is a CEMP-rs star.
The derived atmospheric parameters and abundances are used to infer \HE\ evolutionary status from a comparison with evolutionary tracks.
}
{\HE\ appears to be a giant star below the tip of the red giant branch (RGB). 
The s-process pollution must therefore originate from mass transfer from a companion formerly on the  AGB, now a carbon-oxygen white dwarf (WD). If the 384~d velocity variations are  attributed to the WD companion, its orbit must be seen almost face-on, with $i \sim 2.3^\circ$, because the mass function is very small: $f(M_1,M_2) = (6.1\pm1.1)\times10^{-6}$~\Msun. Alternatively, the WD orbital motion could be responsible for the long-term velocity variations, with a period of several decades. The 384~d variations should then be attributed either to a low-mass inner companion (perhaps a brown dwarf, depending on the orbital inclination), or to stellar pulsations. The latter possibility is made likely by the fact that similar low-amplitude velocity variations, with periods close to 1~yr, have been reported for other CEMP stars in a companion paper (Jorissen et al., 2015). Moreover, Kiss \& Bedding (2003) have shown that Wood's period-luminosity sequence D extends below the RGB tip, corresponding to periods of about 400~d, and is associated with velocity variations.  A definite conclusion about the origin of the 384~d velocity variations should however await the detection of synchronous low-amplitude photometric variations.
}
{}
\keywords{Galaxy: halo -- Stars: AGB and post-AGB -- Stars: carbon -- Stars: evolution -- Stars: individual: HE 0017+0055
}

\maketitle
%

\section{Introduction}

Over the past several years, evidence has been accumulating that as many as 20\% of halo giants with [Fe/H] $<$ -2.5 dex ( a proportion rising to 80\% at [Fe/H] $\le$ -4.0 dex) exhibit overabundances of carbon by as much as 1$-$2 dex \citep[the so-called {\it Carbon-Enhanced Metal-Poor} stars -- CEMP stars;][] {2005ARA&A..43..531B,2010A&A...509A..93M,2014ApJ...797...21P}. Moreover, stars with concomitant enhancement of $s$-process elements (CEMP-s stars) seem to preferentially inhabit the inner halo, while those without $s$-process enhancement (CEMP-no) stars seem to dominate the outer halo \citep{2014ApJ...788..180C}. The conventional explanation for the origin of the CEMP-s stars is mass transfer from a nearby binary companion on the asymptotic giant branch (AGB) stage (which has since evolved into a white dwarf -- WD), by Roche-lobe overflow or by a strong stellar wind, or possibly by a combination of both \citep{2013A&A...552A..26A}, in analogy with barium and CH stars \citep{1990ApJ...352..709M}. Indeed, preliminary radial-velocity data by \citet{2005ApJ...625..825L} suggested that most or all CEMP-s stars are long-period binaries, but the sample then available did not distinguish between CEMP-s and CEMP-no stars, and the results were not conclusive. We have therefore  undertaken similar systematic studies to elucidate the origin of these stars in more detail.  Simultaneously with the present study,  \citet{2015AA....NNN.....H} addressed the same question, with the conclusion that 80\% of CEMP-s 
stars are binaries, and CEMP-no stars are generally not. Besides CEMP-no and CEMP-s stars, there is another class of interest in the context of the present paper, namely the CEMP-rs stars.
In addition to large overabundances of elements produced by the $s$-process, they also exhibit large overabundances of elements traditionally related to the $r$-process. 
The first such stars were discovered by \citet{1997A&A...317L..63B} and \citet{2000A&A...353..557H}. 
 A number (if not all) of these stars exhibit radial-velocity variations
\citep[e.g.,][]{2004A&A...413.1073S,2005A&A...429.1031B,2015AA....NNN.....H}. The nature of the companion star at the origin of the chemical enrichment of the CEMP star, probably through mass transfer, remains however debated \citep{2010A&A...509A..93M,2015AA....NNN.....H}. An intermediate-mass 
AGB star where the $^{22}$Ne($\alpha$,n)$^{25}$Mg neutron source operates is generlly invoked.

\HE\ was included in two independent radial-velocity monitoring programmes, one performed with the HERMES spectrograph attached to the 1.2-m Mercator telescope \citep{2011A&A...526A..69R} and addressing many different science cases \citep{2010MmSAI..81.1022V}; the other with the fibre-fed, bench-mounted spectrograph FIES at the Nordic Optical Telescope (NOT); see \citet{2011ApJ...743L...1H}. Given the unusual properties of \HE, we decided to combine our data sets and discuss that star separately from the remainder of the samples \citep[see][for a discussion of the results of the HERMES monitoring of CEMP-s stars, and Hansen et al. 2015b, for a discussion of the NOT results]{2015A&A...XXX..NNNJ}.

\HE\ was discovered in a survey by \citet{1985AJ.....90..784S} with a spectrum resembling type R, 
and received number 39 in the second edition of the {\it General catalog of Galactic carbon stars} 
\citep[GCGCS;][]{1989PW&SO...3...53S}. It was later rediscovered in the framework of the Hamburg / ESO survey
\citep{2001A&A...375..366C} as a high-latitude carbon star ($b = -61^{\circ}$).  Based on a fit of selected Ca and Fe lines 
\citep[see][]{2007AJ....133.1193B} with synthetic spectra, \citet{2011AJ....141..102K} derived a metallicity of $-2.7$ for \HE . 
Hence the star belongs to the family of CEMP stars. 

The very low surface gravity ($\log g = 0.18$) adopted by \citet{2011AJ....141..102K} for \HE\ would make  it   
a very luminous CEMP star, a possible twin of \CS\ \citep{2006A&A...455.1059M} and HD~112869 \citep{2015ApJ...803...17Z}. Given the scarcity of  such low-metallicity AGB stars, finding yet another example appears surprising and warrants the detailed study presented in this paper. In fact,  based on the  stellar parameters derived in  Sects.~\ref{Sect:parameters} and \ref{Sect:sprocess},
our study does {\it not} confirm, however, the AGB nature of \HE, as we discuss in Sect.~\ref{Sect:evol}.  
But we start the paper by presenting the radial-velocity monitoring and the resulting orbital parameters (Sect.~\ref{Sect:orbit}). 

\section{Radial-velocity monitoring and orbital solution}
\label{Sect:orbit}

\subsection{Radial-velocity monitoring with the HERMES spectrograph}
\label{Sect:HERMES} 

The HERMES radial-velocity (RV) monitoring was performed with the fibre-fed HERMES spectrograph attached to the 1.2-m Mercator telescope of the Katholieke Universiteit Leuven, installed at the Roque de los Muchachos Observatory (La Palma, Spain) and fully described in \citet{2011A&A...526A..69R}. HERMES is designed to 
optimise both stability and efficiency and covers the whole wavelength range from 
380 to 900~nm at a resolving power of $\sim86\,000$. The high-resolution science fibre has a 2.5 arcsec aperture on the sky, using a two-slice image 
slicer. 

\begin{figure}
\includegraphics[width=9cm]{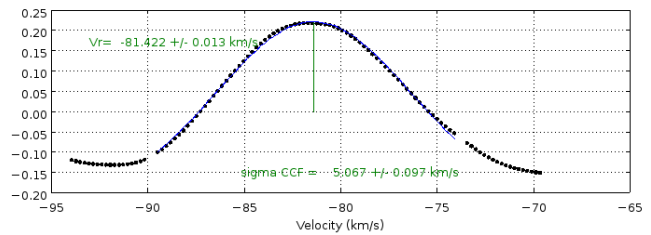}
\caption{\label{Fig:CCF}
A typical CCF for \HE\ from a spectrum taken with an exposure time of 900 sec. The vertical line shows the radial velocity of the star as derived from a Gaussian fit to the CCF core. }
\end{figure}

A Python-based pipeline extracts a wavelength-calibrated and a
cosmic-ray cleaned spectrum. A separate routine is used for measuring
RVs, by means of a cross-correlation with a spectral mask constructed
from a carbon-star spectrum.  A restricted region, covering the range
478.11 -- 653.56~nm (orders 55 -- 74) and containing 2103 useful
spectral lines, was used to derive the RV, in order to avoid telluric
lines on the red end and crowded and poorly-exposed spectra on the blue end. A
spectrum with a signal-to-noise ratio of 15 is usually sufficient to
obtain a cross-correlation function (CCF) with a well-pronounced
maximum.  A typical CCF is shown in Fig.~\ref{Fig:CCF} for a spectrum with a SNR of 15 in the selected spectral region. 

Radial velocities are determined from a Gaussian fit to the core of the CCF 
with an internal precision of a few m/s. The most important external source of error is the varying atmospheric pressure in the spectrograph room \citep[see Fig.~9 of][]{2011A&A...526A..69R}, which is largely eliminated by the arc spectra taken 
for wavelength calibration. The long-term stability (years) of the resulting radial velocities is checked with RV standard stars from \citet{1999ASPC..185..367U}. 
Their standard-deviation  distribution peaks at $\sigma(Vr)=55$~m/s \citep[see also Fig.~2 of ][]{2015A&A...XXX..NNNJ}, which we adopt as the typical radial-velocity uncertainty for such relatively bright stars. The RV standard stars have also been used to tie the HERMES RVs to the IAU standard system.

The list of HERMES velocities is given in Table~\ref{Tab:HERMES_Vr}.

\setlength{\tabcolsep}{5pt}
\begin{table}
\caption{\label{Tab:HERMES_Vr}
HERMES/Mercator and FIES/NOT radial velocities of  \HE\ (without any zero-point correction). 
The uncertainty quoted on the FIES RVs corresponds to the standard deviation of the correlation results from all spectral orders. 
 For the HERMES data, the uncertainty is instead the formal error on the CCF Gaussian fit on 20 orders. A more realistic uncertainty is the long-term error describing the long-term stability of the spectrographs (see text).
}

\begin{tabular}{lllllllllll}
\hline\\
\multicolumn{1}{c}{JD} & \multicolumn{2}{c}{$Vr\pm\epsilon$} & \multicolumn{1}{c}{JD} & \multicolumn{2}{c}{$Vr\pm\epsilon$}\\
(-2\,400\,000) & \multicolumn{2}{c}{(\kms )} & (-2\,400\,000) & \multicolumn{2}{c}{(\kms ) }\\
\hline
\medskip\\
\multicolumn{3}{c}{HERMES} & \multicolumn{3}{c}{NOT}\\
\medskip\\
 55053.5756 &-79.423&0.008&  54314.6702  &-78.622  &0.007\\
 55053.5890 &-79.431&0.005&  54338.6419  &-78.944  &0.007\\
 55053.6025 &-79.382&0.009&  54373.6224  &-78.691  &0.009\\
 55087.5807 &-79.899&0.011&  54396.5371  &-78.583  &0.007\\
 55087.5939 &-79.823&0.006&  54406.5966  &-78.505  &0.012\\
 55087.6072 &-79.851&0.006&  54480.3868  &-78.183  &0.012\\
 55098.6838 &-79.597&0.008&  54793.4846  &-79.098  &0.018\\
 55098.6976 &-79.590&0.006&  54820.3386  &-78.830  &0.017\\
 55098.7124 &-79.738&0.007&  55059.7365  &-79.922  &0.008\\
 55159.5055 &-79.417&0.006&  55149.4730  &-79.647  &0.010\\
 55159.5201 &-79.429&0.006&  55207.3499  &-79.391  &0.015\\
 55159.5343 &-79.463&0.014&  55415.6081  &-80.439  &0.009\\
 55418.6223 &-80.099&0.008&  55439.5914  &-80.778  &0.010\\
 55423.7016 &-80.118&0.003&  55503.4086  &-80.330  &0.023\\
 55497.4806 &-79.992&0.007&  55738.7344  &-80.148  &0.008\\
 55938.3316 &-79.977&0.008&  55776.6821  &-80.675  &0.008\\
 55956.3373 &-79.944&0.003&  55821.5766  &-80.870  &0.009\\
 56109.7113 &-80.056&0.005&  55944.3253  &-80.429  &0.011\\
 56140.7386 &-80.201&0.006&  56139.7129  &-80.857  &0.010\\
 56311.3363 &-80.588&0.007&  56241.3910  &-81.420  &0.018\\
 56481.7027 &-80.278&0.007&  56545.6267  &-81.474  &0.011\\
 56511.7176 &-80.560&0.007&  56652.4180  &-81.951  &0.015\\
 56557.6467 &-80.868&0.006&  56686.3209  &-81.538  &0.015\\
 56634.3763 &-81.416&0.007&  56840.7182  &-80.866  &0.028\\
 56849.6608 &-80.395&0.017&  56888.5421  &-81.261  &0.013\\
 56904.6038 &-80.550&0.016&  56917.5981  &-81.398  &0.014\\
 56947.5245 &-80.919&0.014&  56987.3813  &-81.885  &0.009\\
 56994.3660 &-81.219&0.013 & 57257.5774  &-81.396  &0.010\\
 56996.3424 &-81.103&0.012\\
 57052.3396 &-81.528&0.005\\
 57052.3506 &-81.422&0.013\\
 57056.3370 &-81.287&0.018\\
 57056.3498 &-81.274&0.017\\
\medskip\\
\hline
\end{tabular}
\end{table}

\subsection{Radial-velocity monitoring at the NOT}
\label{Sect:NOT} 

The NOT spectra were obtained in service mode at roughly monthly intervals with the FIES spectrograph\footnote{http://www.not.iac.es/instruments/fies/}, which covers the fixed wavelength range 364\,{nm} $-$ 736\,{nm} in 78 orders at a resolving power of $R\sim46\,000$. The average SNR of the spectra is $\sim$10, ranging from $\sim$2 to $\sim$20, and is obtained in $\sim$20~min on a star of $V = 14.5$. Integrations of 15 min or longer are split into three in order to enable effective cosmic-ray rejection, and a Th-Ar reference spectrum is observed immediately before every stellar spectrum. $2-3$ RV standard stars from a fixed set of seven, also selected from \citet{1999ASPC..185..367U}, were observed on every observing night in order to monitor any zero-point variations in the derived velocities. 

The observations were reduced and correlated with software developed by Lars Buchhave, originally to deliver high-precision radial velocities of exo-planet host stars from echelle spectrographs, especially from FIES \citep{2010ApJ...720.1118B}. Using the highest-signal spectrum of \HE\ itself as the template spectrum, multi-order cross-correlation and optimised filtering is performed, and the final radial velocity, given in Table~\ref{Tab:HERMES_Vr}, is determined by a Gaussian fit to the peak of the CCF. The internal error of each velocity, also listed, is the standard deviation of the results from all spectral orders. Correlation with a ``Delta'' template, a synthetic spectrum consisting of $\delta$ functions at the (solar) wavelengths of selected lines, is used to determine the radial velocity of the \HE\ template and thus that of the whole data set. More details are given in \citet{2015AA....NNN.....H}, from where 
the NOT data in Table~\ref{Tab:HERMES_Vr} are taken.  

The observations of the (much brighter) RV standard stars were reduced exactly as for all science targets. The typical standard deviation is 38 m~s$^{-1}$ for a star with $\sim$41 observations over $\sim$2700 days, or 7.5 years, and the mean differences between the NOT and \citet{1999ASPC..185..367U} radial velocities is $-13 \pm38$ m~s$^{-1}$. 

The velocity curve from the combined HERMES+NOT data set covers 2943 days or 8 years with a mean radial velocity of $\sim -80$ \kms. It is presented in Fig.~\ref{Fig:Vr}, which clearly shows 8 cycles of a short-term periodic velocity oscillation, superimposed on a long-term trend. We have found that the latter is best modelled by a quadratic polynomial of the form $Vr = 3.252\times 10^{-7} x^2  -0.0374 x + 993.215$, where $x = JD - 2400000$, while the short-term oscillation is adequately represented by a Keplerian orbit. 

\subsection{Orbital solution}

\begin{figure}
\includegraphics[width=9cm]{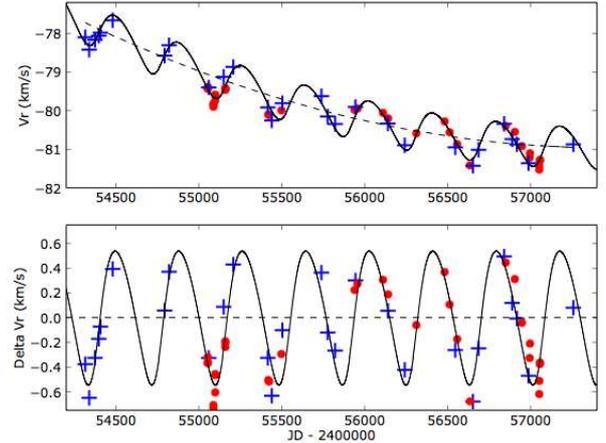}
\caption{\label{Fig:Vr}
The radial-velocity curve of \HE\ showing the short- and long-term orbital fits combined (top) as well as the short-term orbit alone (red dots: HERMES velocities; blue plusses: NOT velocities).
}
\end{figure}

From trial orbital solutions of the separate HERMES and NOT, we found that adding a small zero-point correction of 0.52~\kms\ to the NOT velocities would minimise the standard deviation of the merged data set in the combined orbital solution, the elements of which are given in Table~\ref{Tab:orbit}. Fig.~\ref{Fig:Vr} shows the 
velocity curve from the complete HERMES+NOT data set as well as the short-term orbit alone.

\begin{table}
\caption{\label{Tab:orbit}
Orbital elements of the presumed short-period pair.
}
\begin{tabular}{lll}
\hline\\
$P$ (d) & $384.6\pm0.9$\\
$e$       & $0.15\pm0.04$\\
$\omega$ ($^{\circ}$) & $234\pm16$\\
$T_0$   &$2455908.0 \pm 15.7$ \\
$V_0$ (\kms) & $0.04\pm0.02$ \\
$K$ (\kms) & $0.54\pm0.03$\\
$f(M_1,M_2)$ (\Msun) & $(6.1\pm1.1)\times10^{-6}$\\
$a_1 \sin i$ (Gm) & $2.8\pm0.2$ \\
$\sigma$ (\kms) & 0.14\\ 
$N$ & 61\\
\hline\\
\end{tabular}

 Note. The table lists the orbital parameters, as follows: $P$ is the orbital period; $e$ is the orbital eccentricity; $\omega$ is the argument of periastron; $T_0$ is the time of passage at periastron; $V_0$ is the velocity of the barycentre; $K$ is the semi-amplitude of the orbit: $f(M_1,M_2)$ is the mass function; $a_1$ is the semi-major axis of the orbit of the visible component around the centre of mass of the system; $i$ is the inclination of the orbital plane with respect to the plane of the sky; $\sigma$ is the rms of the $O-C$ residuals; $N$ is the number of data points.
\end{table}

With the highly significant radial-velocity amplitude and stable period over eight orbital cycles, 
and in absence of any photometric evidence for stable pulsations  (this aspect will be further discussed in Sect.~\ref{Sect:photometry}  in relation with Fig.~\ref{Fig:ASAS}), it seems natural to interpret our velocity data as implying that \HE\ is a triple star system with an inner orbit of low eccentricity (which is either seen nearly face-on, or with a low-mass companion) and a more distant companion. These alternatives will be discussed in the next section (Sect.~\ref{Sect:binary}). 
Note that the shorter period is sufficiently different from 365~d that phases have shifted by about half a cycle over the 8-year span of our monitoring, relative to any residual seasonal effects, and we have documented elsewhere \citep{2011ApJ...743L...1H} that we can reliably measure binary motions of similar period and even lower amplitude. The outer orbit must have a period of several decades, since we only see marginal evidence of a gradual acceleration over $\sim3000$~d.

Considering the fact that \HE\ is a CEMP-rs star (as we will show in Sect.~\ref{Sect:sprocess}), not residing on the AGB (see Sect.~\ref{Sect:evol} and Fig.~\ref{Fig:HR}), it must owe its peculiar abundances to mass transfer in a binary system, like barium and CH stars \citep{1990ApJ...352..709M,2013EAS....64..163G,2015A&A...XXX..NNNJ}. The most recent collection of orbital elements for these families, presented in the 
two recent references above, indicates that their orbital periods range between 80~d and up to 50~yr or more.  Both the inner and outer companions of \HE\ have periods in this range, consistent with that of barium and CH systems. This argument based on orbital periods cannot be used thus to tag which, among the inner or outer companion, is the WD responsible for the chemical pollution. 
In case the short-period variations turn out not to be of Keplerian origin (as we will argue in Sect.~\ref{Sect:photometry}), the long-term velocity variations must then be attributed to the WD companion. As discussed above, this poses no problem in terms of accretion efficiency, since orbital periods of several decades are known among  CH and barium stars.

\subsection{Origin of the short-term velocity variations}  
\label{Sect:binary}

The remarkable feature of the short-term orbit presented in Table~\ref{Tab:orbit} is the very small value of its mass function [$(6.1\pm1.1)\times10^{-6}$~\Msun]. 
In this section, we discuss three possible interpretations of this fact: (i) an orbit with a very low inclination; (ii) a companion of very low mass; (iii) a spurious Keplerian solution, the velocity variations being caused by envelope pulsations.

\subsubsection{An orbit with a very low inclination?}

The small mass function derived from the orbital fit could imply a very low orbital inclination. Assuming $M_1 = 0.9$~\Msun\ for the primary component of \HE, $i \sim 2.3\degr$ is required to yield a secondary mass of $M_2 = 0.6$~\Msun, typical for a CO WD. In this situation, it is thus the inner companion which is responsible for the chemical pollution, whereas the outer companion is an innocent bystander. This situation immediately implies that the two orbits cannot be co-planar. If they were so, the semi-amplitude of the outer orbit (assumed to be 3~\kms; Fig.~\ref{Fig:Vr}), the inferred inclination ($2.3\degr$) and period ($\sim 4\times8 = 32$~yr) would imply a mass of several $10^3$~\Msun\  for the outer companion, which is of course impossible.
Indeed, many non-coplanar triple systems are known \citep{2003AJ....125.2630H,2008AJ....135..766M,2011ApJ...728..111O}, and modern theories of stellar
formation \citep{2002A&A...384.1030S,2007ApJ...669.1298F} predict that triple systems will form non-coplanarily.

\subsubsection{An orbit with a very low-mass companion?}

An alternative solution is that of a very low-mass companion. If $M_1 = 0.9$~\Msun\ as assumed above, then the derived mass function imposes that the companion be more massive than 0.017~\Msun.  A brown-dwarf companion is expected for orbital inclinations in the range $12.4^\circ \le i \le 90^\circ$, $i = 12.4^\circ$ corresponding to the brown-dwarf threshold at $M_2 = 0.079$~\Msun.
Such a brown-dwarf companion, if confirmed, would be the most metal-poor ([Fe/H]$ \sim -2.4$ as shown  in Sect.~\ref{Sect:re-analysis}) ever discovered, since so far brown-dwarf metallicities are all above [Fe/H]$ \sim -0.6$ \citep[][and references therein]{2014MNRAS.440..359B}, with one possible exception \citep[with a metallicity somewhere between $-1$ and $-2$;][]{2003ApJ...592.1186B}.

\subsubsection{Envelope pulsation causing spurious Keplerian-like velocity variations?}
\label{Sect:photometry}

Although the Keplerian interpretations  presented above are in principle perfectly acceptable,  the study of low-metallicity giants by \citet{2003AJ....125..293C} and  by \citet{2015A&A...XXX..NNNJ}  in a companion paper invites us, however,  to be cautious.

Among 13 CEMP-(r)s and CH binary systems, \citet{2015A&A...XXX..NNNJ} have identified two more cases similar to \HE, namely HE~1120-2122 and HD~76396,  where small-amplitude variations ($K$ ranging from 0.1 to 0.9~\kms) with periods very close to 1~yr are superimposed on a long-period Keplerian orbit. The CEMP-s star CS~30322-023 shows as well a velocity curve similar to that of \HE, with short-term radial-velocity variations with a period of 192~d, and a semi-amplitude of 1.7~km~s$^{-1}$, possibly superimposed on a long-term trend \citep[see Fig.~7 of][]{2006A&A...455.1059M}.

On the other hand, \citet{2003AJ....125..293C} reviewed previous evidence that low-metallicity giant stars, especially in globular clusters, systematically exhibit radial-velocity jitter for luminosities near or above the RGB tip (i.e., $M_V \le -1.5$ or $\log g \le 1.3$). A closer look at these velocity variations 
reveals that they mimick Keplerian variations, with semi-amplitudes of the order of  1 -- 1.5~\kms\ and periods of the order of 170 - 190~d. Among the cases reported above for CEMP-(r)s/CH systems, only CS~30322-023 conforms to those properties, the other systems being of lower amplitude and longer period. It is likely, however, that those were out of reach with the spectrograph used by  \citet{2003AJ....125..293C}.

An important property of  the velocity variations reported by \citet{2003AJ....125..293C} is that they are correlated with low-amplitude photometric variations: {\it Hipparcos} observations of HD~3008 and HD~110281 in Carney's sample for example revealed them to be variables with (peak-to-peak) amplitudes $A = 0.134\pm0.045$~mag and $A = 0.195\pm0.032$~mag, respectively. Most importantly, the photometric and radial-velocity data appear to vary with the same period. 

\begin{figure}
  \centering \includegraphics[width=8cm]{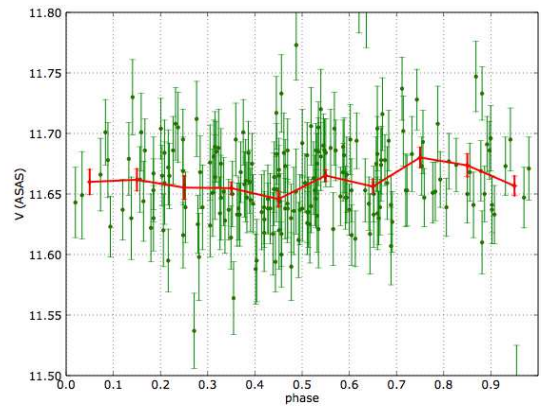}
    \caption{Photometric data ('grade A' only) from the ASAS-3 catalogue for \HE\ = ASAS 002022+0112.1, folded on the 383~d period inferred from the radial-velocity variations. The span of the ordinate axis has been set in such a way to eliminate outlying data points. The red curve with error bars corresponds to the average values of the $V$ magnitude in bins of width 0.1 in phase, after eliminating outlying points (i.e., those outside the graphical window). The error bars along that curve correspond to the error on the mean, i.e., the standard dispersion in the considered bin divided by $\sqrt{N}$, where $N$ is the number of data points in the bin.
    }
  \label{Fig:ASAS}
\end{figure}

Photometric data for \HE\ are available from the ASAS-3 survey \citep{1997AcA....47..467P}, and are displayed in 
Fig.~\ref{Fig:ASAS}, folded on the radial-velocity period of 384~d. No significant modulation has been detected at a 3$\sigma$ level of  0.03~mag (where $\sigma = 0.01$~mag is the typical error bar on the light curve in  Fig.~\ref{Fig:ASAS}). Thus, one may exclude (sinusoidal)  variations with a peak-to-peak amplitude larger than $0.03 \times 2\sqrt{2} = 0.08$~mag in \HE. 

Should the radial-velocity variations be due to stellar oscillations,
they would not contradict the absence of ligh-variability detection. Indeed, 
from a simple linear theory of  oscillations, \citet{1995A&A...293...87K}  relate 
the standard deviation $\sigma_\lambda$ (in millimag) of the light variations at wavelength $\lambda$ (in nm) 
to the radial-velocity standard deviation $\sigma ({\mathrm V_r})$ (in \kms) caused by adiabatic acoustic oscillations. They find the relation:
\begin{equation}
\frac{\sigma_\lambda}{\sigma({\mathrm V_r})} = \frac{21.8}{(\lambda/550) (T_{\rm eff}/5777)^2}.
\end{equation}
Considering the 0.6~\kms\  semi-amplitude $K$ of the velocity variations which turns into $\sigma({\mathrm V_r}) = K / \sqrt{2} = 0.42$~\kms, 
the above relation predicts   $\sigma_V = 18$~millimag, which is smaller than the detection threshold for the ASAS data
estimated above. Thus, the hypothesis that the short-term velocity variations are due to stellar oscillations is
 consistent with the absence of large light variations in the ASAS light curve.

The expected level of photometric variability in giant stars just below the tip of the red giant branch, where \HE\ seems to be located (Sect.~\ref{Sect:evol}), has been investigated as well by \citet{2003MNRAS.343L..79K} for the Large Magellanic Cloud. These authors conclude that most of these stars are variable, but 
with variability (peak-to-peak) amplitudes generally smaller than 0.14~mag.
Remember that  the ASAS-3 survey puts an upper limit on the (peak-to-peak) variability amplitude of \HE\ at 0.08~mag.

In the absence of more accurate photometric data for \HE, we cannot be sure that the short-term radial-velocity variations are associated to some kind of envelope pulsations. However, the accumulation of systems among CEMP-(r)s/CH stars with low-amplitude velocity variations on a time scale of 190 to 380~d is puzzling 
and invites us to be cautious, but at this stage we should not exclude any of the three possibilities described in this and the previous sections.

\section{Stellar parameters}
\label{Sect:parameters}

We now turn to a description of the atmospheric parameters of \HE, necessary to perform the abundance analysis which will reveal the CEMP-rs nature of \HE. But we present first a description of its kinematical properties, because these properties will guide the derivation of the star's surface gravity.

\subsection{Kinematics}
\label{Sect:kinematics}

 The Tycho-2 project \citep{2000A&A...355L..27H} derived 
the proper-motion values 
$\mu_{\alpha}\;\cos\delta = 6.9\pm2.7$~mas~yr$^{-1}$ and $\mu_{\delta} =  -14.2\pm2.5$~mas~yr$^{-1}$. These 
proper motions (along with the radial velocity of $-80$~\kms; see Table~\ref{Tab:HERMES_Vr}) have been converted into space velocities $U,V,W$, as this will be the basis for an important argument in the choice of the atmospheric parameters. 
$U,V,W$  are the heliocentric velocity components along the Galactic coordinate system ($U$ is positive in the direction of the Galactic centre, and $V$ is positive in the direction of Galactic rotation $l = +90^{\circ}$). 
Adopting different possible locations for \HE\ in the Hertzsprung-Russell diagram (on the asymptotic giant branch, on the giant branch, or on the main sequence, along the evolutionary track of a 0.9~\Msun\ star of metallicity [Fe/H] = -2.4; Table~\ref{Tab:space}), the corresponding luminosities yield  distances that are 
used to convert the Tycho-2 proper motions into components of the space velocity. In this way, we link the atmospheric gravity $\log g$ with the space velocity.

\begin{table*}
\caption[]{\label{Tab:space}
Kinematical properties of \HE, for different hypotheses about its absolute visual magnitude $M_V$ and hence distance, adopting the Tycho-2 proper motion and a $V$ magnitude of  11.65 \citep[][in agreement with the ASAS-3 data; see Fig.~ \protect\ref{Fig:ASAS}]{2000A&A...355L..27H}. The uncertainties on the space motion originate from the uncertainties on the Tycho-2 proper motion. The correspondence between $M_V$ and $\log g$ 
is from the STAREVOL track (see Sect.~\ref{Sect:evol}) for a 0.9~\Msun\ star with [Fe/H] = $-2.4$, [C/H] = $-0.7$, and C/O = 20, adopting a bolometric correction of 0.3. The first line corresponds however to the Padova track used by \citet{2011AJ....141..102K}.  $U,V,W$  are the heliocentric velocity components along the Galactic coordinate system ($U$ is positive in the direction of the Galactic centre, and $V$ is positive in the direction of Galactic rotation $l = +90^{\circ}$). $V_{\alpha*}$ and $V_{\delta}$  are the space velocities along the direction of right ascension and declination, respectively.
}
\begin{tabular}{rlrrccccccccl}
\hline\\
\Teff & $\log g $ & \multicolumn{1}{c}{$L$} & $M_V$ & \multicolumn{1}{c}{$d$} & $|z|$ & $V_{\alpha*}$ & $V_{\delta}$ & \multicolumn{1}{c}{$U$} & \multicolumn{1}{c}{$V$} & \multicolumn{1}{c}{$W$} &lum. class & model\\
(K)    &               & (L$_{\odot}$)  &          & \multicolumn{1}{c}{(kpc)} & (kpc) & (\kms)    & (\kms) & (\kms) & (\kms) & (\kms) \\
\hline\\
4185 & 0.18  & 4470  & $-4.1$ & 14 & 12.2 & $458\pm179$ & $942\pm166$ & 68  &    -956    &  -433 & II & PADOVA\\
3950 & 0.5    &1920   & $-3.2$ & 9.3 & 8.1 & $305\pm119$ & $628\pm111$ & 49  &    -647    &  -264 &II & STAREVOL\\
4000 & 1.0    &  523    & $-1.8$ & 4.8 & 4.2  &$157\pm69$ & $296\pm64$ & 29    &  -348    &  -110 &III& STAREVOL\\
3950 &  4.9   & 0.03   &  8.8  & 0.037& 0.032 & $1.2\pm0.5$  & $0.15\pm0.03$ & 12  &    -40  &     68 & V& STAREVOL\\
\hline\\
\end{tabular}
\end{table*}

The results are listed in Table~\ref{Tab:space} and reveal that for gravities lower than $\log g = 0.5$, the space velocity is uncomfortably large, since it is likely to exceed the Galactic escape velocity\footnote{See \protect\citet{2012A&A...543A..58P} for a case study of high-velocity CH stars (HD~5223 and BD~-32$^{\circ}1346$), near
or above the Galactic escape velocity.}. 
In the solar neighbourhood, the Galactic escape velocity is estimated to be about 500 -- 600~\kms\ \citep{2007MNRAS.379..755S}, but it must be smaller in the halo at the distance of \HE\ (we will show later that the star is unlikely to be a main-sequence star; hence, it must be located at a large Galactic scale height). 
The $U,V,W$ velocities obtained for $\log g = 1$ (Table~\ref{Tab:space}) are consistent with a typical halo population, since $V \sim -350$~\kms\ combined with its estimated metallicity [Fe/H]$ = -2.4$ (see below) locates \HE\ right along the trend observed by \citet{1996AJ....112..668C} for halo stars (see their Figs.~1, 3, 5, and 8).

\subsection{Atmospheric parameters: previous studies}
\label{Sect:previous}

\citet{2011AJ....141..102K} have derived atmospheric parameters for \HE: the effective temperature 
\Teff~=~4185~K has been estimated from the  (\Teff , $V-K$) relationship  of  \citet{1996A&A...313..873A},
with $V-K = 2.96$, $K$ being adopted from the 2MASS catalogue \citep[$J = 9.31\pm0.03$, $H= 8.70\pm0.03$, 
$K = 8.50\pm0.02$;][]{2003yCat.2246....0C}, and $V = 11.459\pm0.008$ and $B = 12.991\pm0.013$ from 
\citet{2007ApJS..168..128B}, yielding $B - V = 1.532$. 

\citet{2011AJ....141..102K} estimated the surface gravity 
to be $\log g = 0.18$, because that value matches \Teff~=~4185~K 
on the Padova evolutionary tracks for the inferred metallicity of [Fe/H]$= -2.5$  
\citep{2000A&AS..141..371G,2001A&A...371..152M}, {\it thus assuming that the star lies on the giant branch} and neglecting the possibility that it could be a dwarf carbon star. 
Uncertainties in \Teff\ and $\log g$ are 100~K and 0.5~dex, 
respectively. The microturbulence is taken to be 2~km~s$^{-1}$, consistent with previously determined 
microturbulence values for giant CEMP stars 
\citep{2007ApJ...658.1203J,2007ApJ...655..492A}. 
We showed in Sect.~\ref{Sect:kinematics} 
that a low gravity yields an unrealistic value for the space motion of \HE\ (since the inferred high luminosity puts the star at a distance of 14~kpc; Table~\ref{Tab:space}), and warrants a revised analysis
of the atmospheric parameters as presented in Sect.~\ref{Sect:re-analysis}.

Adopting nevertheless the above atmospheric parameters,  \citet{2011AJ....141..102K}  estimated a metallicity [Fe/H]$ = -2.72$ for \HE , based on a fit of selected Ca and Fe lines with synthetic spectra
\citep[see][]{2007AJ....133.1193B}. \citet{2011AJ....141..102K}  
also derived [C/H]$ = -0.41$, [C/Fe]$ = 2.31$, [N/Fe]$ = 0.52$, [O/H]$ = -1.7$,  and [O/Fe]$ = 1.0$ \citep[corresponding to $\log\epsilon(\mathrm{C}) = 8.02$, $\log\epsilon (\mathrm{N}) = 5.64$, $\log\epsilon (\mathrm{O}) = 6.99$, adopting $\log\epsilon_{\odot} (\mathrm{C}) = 8.43$,
$\log\epsilon_{\odot} (\mathrm{N}) = 7.83$, and $\log\epsilon_{\odot} (\mathrm{O}) = 8.69$ for the Sun;][]{2009ARA&A..47..481A}.  In this paper, we use the standard notation $\log\epsilon(X) = \log [n(X)/n(H)] + 12$, where $n(X)$ is the number density of element X.
The CNO abundances of \HE\ appear normal among CEMP stars \citep[see, e.g., Fig.~18 of][]{2010A&A...509A..93M}.
Adopting the solar CNO abundances as listed above [corresponding to (C/O)$_{\odot} = 0.55$], one thus finds C/O$ = 10.7$ for \HE. 
\citet{2005MNRAS.359..531G} moreover finds a very low $^{12}$C/$^{13}$C ratio of 1.3, typical of (or even lower than) the 
CN-cycle equilibrium value. This value is the lowest among those reported by both \citet{2005MNRAS.359..531G} and   
\citet{2010A&A...509A..93M}.

Nothing is known about the s-process content of \HE ; this issue will therefore be investigated as well in 
Sect.~\ref{Sect:sprocess}.

\subsection{Atmospheric parameters revisited}
\label{Sect:re-analysis}

We re-derived the atmospheric parameters of \HE\ for the reasons given in Sect.~\ref{Sect:previous}. 
We first tried to use the relation between $(V - K)$, $(H - K)$, $(J - K)$  and \Teff\ derived by \citet{2001A&A...369..178B} for carbon stars to estimate the effective temperature.  Unfortunately, the photometric indices for \HE\ ($V - K = 2.96\pm0.03$, $H - K = 0.20\pm0.05$, $J - K = 0.81\pm0.05$) 
require to severely extrapolate the calibrations of \citet{2001A&A...369..178B}, which were aimed at cool carbon stars of type N. Taken at face value, these calibrations yield temperatures of 4680, 3470, and 3890~K, respectively, and are thus of little use. We note, however, that \HE\ has colours identical to those of 
the warm R  stars studied by \citet{2009A&A...508..909Z} (see their Fig.~1). In that sample, HIP~87603 and HIP~113150 share the same 
 $H - K$ and $J - K$ indices as \HE, and have effective temperatures of 4100 and 4500~K respectively (and both with $\log g = 2.0$). 

The photometric similarity between \HE\ and both R and CH stars\footnote{We note, however, that given the strong CH band at 437.0~nm present in \HE, 
it is clear that the star must be considered as a CH star \citep[see Fig.~1e of][]{1996ApJS..105..419B}. The atmospheric parameters of CH and R stars are anyway very similar.} \citep{2009A&A...508..909Z} led us to focus our search around \Teff \ $\sim$ 4100 -- 4500~K and $\log g = 2.0$. The final atmospheric parameters were derived by an iterative process of fitting specific spectral regions and fine-tuning the parameters consecutively, modifying \Teff, gravity, metallicity, s-process abundances as a 
whole, and
the CNO abundances until a reasonable fit to the observed spectrum was found, for the whole range 420 -- 800~nm. The synthetic spectra were produced by the TURBOSPECTRUM code \citep{1998A&A...330.1109A} from carbon-enhanced model atmospheres with the C, N and O abundances iteratively adapted to the values found in \HE . The model atmospheres were computed with the opacity sampling technique usual for the MARCS models \citep{2008A&A...486..951G}, and using opacity tables with variable C, N and O abundances. 

\begin{figure}
\includegraphics[width=9.5cm]{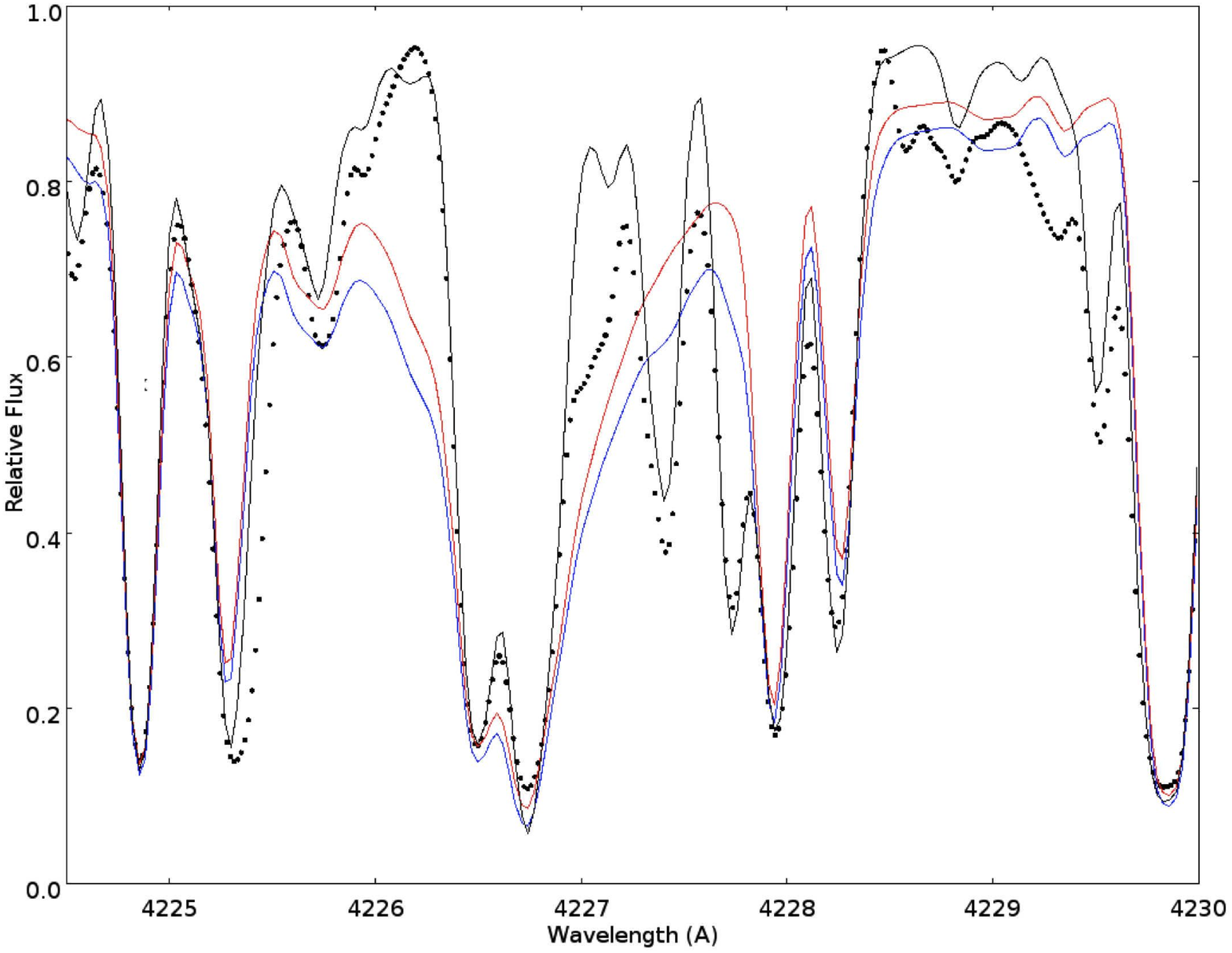}
\caption{\label{Fig:Ca}
Effect of gravity on the \ion{Ca}{I} $\lambda$~422.67~nm line (black curve: $\log g= 1$ and [Fe/H] = $-2.4$~dex; blue curve: $\log g = 5$ and [Fe/H] = $-2.4$~dex; red curve: $\log g= 5$ and [Fe/H] = $-3.5$~dex).  Except otherwise stated, the atmospheric parameters used to generate the synthetic spectra are those listed in Table~\ref{Tab:parameters}.
}
\end{figure}

\begin{figure*}
\includegraphics[width=20cm]{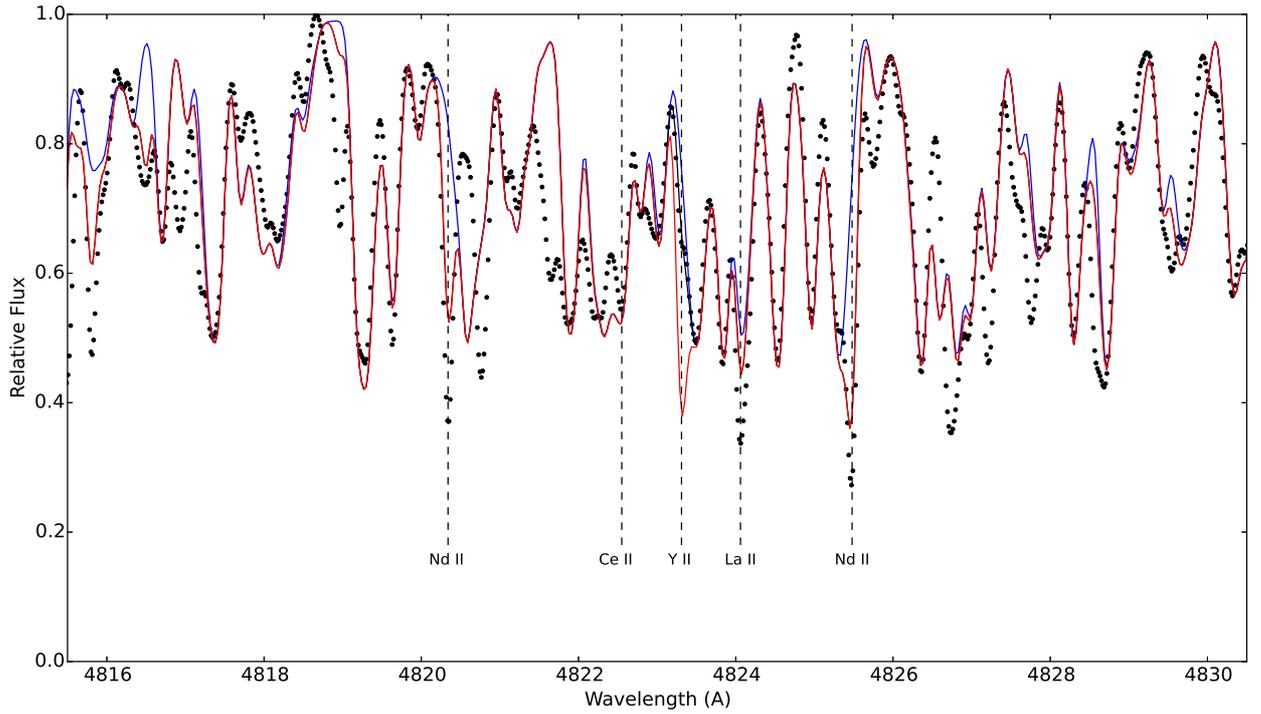}
\caption{\label{Fig:Ba}
The effect of a general increase  by 2.0~dex of s-process elements (red line), as compared to their solar abundances (blue line). 
Compare with Fig.~5 of \citet{2007A&A...461..641R}. Note that the \ion{Y}{II} line around 4823.3~\AA\ is not as enhanced as the other s-process lines (red line).
 A better match to the Y abundance is obtained with [Y/Fe] = 0.5~dex (black line).
Except otherwise stated, the atmospheric parameters used to generate the synthetic spectra are those listed in Table~\ref{Tab:parameters}. 
}
\end{figure*}

\begin{figure*}
\includegraphics[width=20cm]{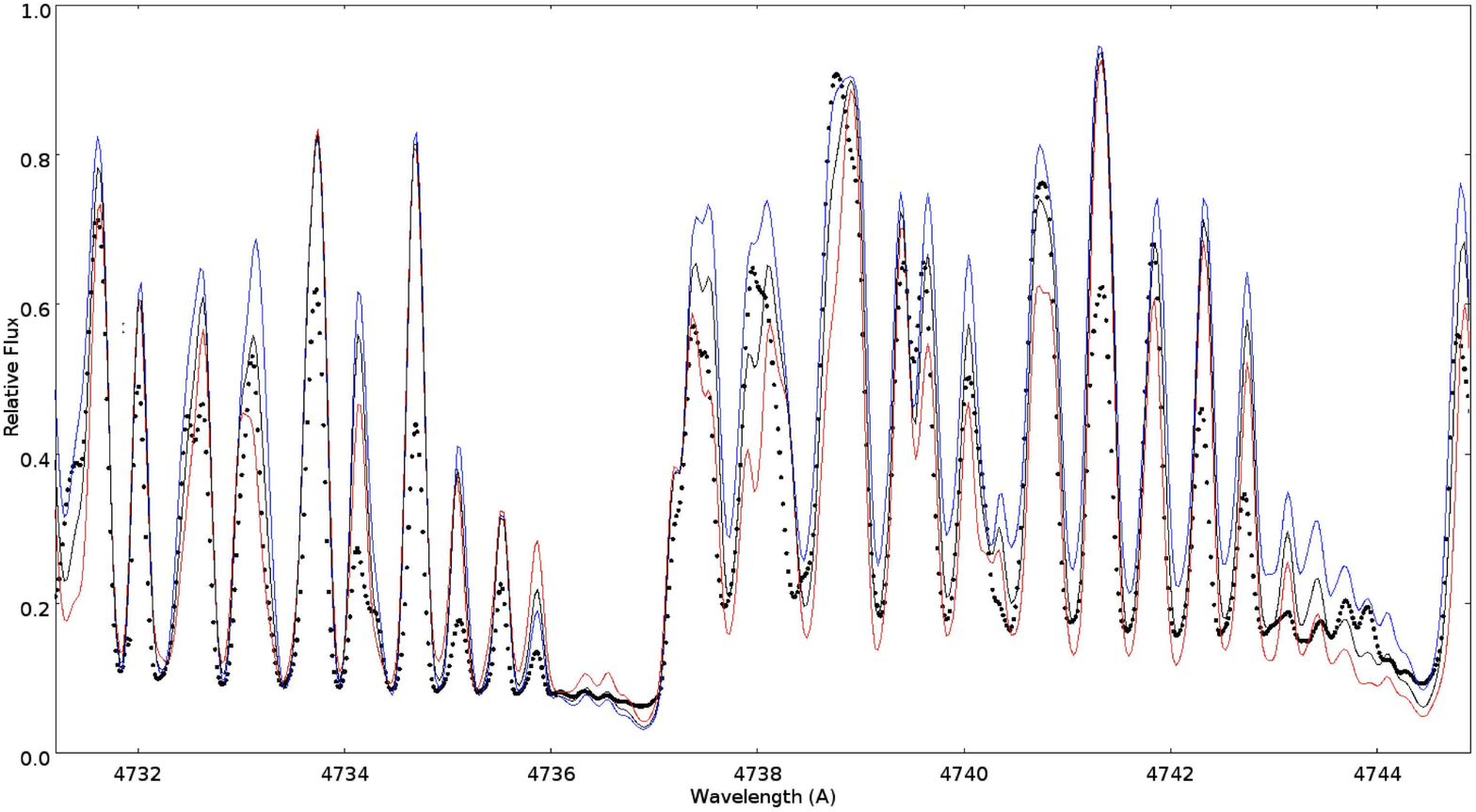}
\caption{\label{Fig:C12C13}
The $^{12}$C$^{13}$C and  $^{13}$C$^{13}$C band heads around 474.0 nm (with $\log\epsilon$(C) = 8.2 and   $\log\epsilon$(N) = 7.9), for  $^{12}$C/$^{13}$C = 2
(red curve), 4
(black curve), and 7 (blue curve). Except otherwise stated, the atmospheric parameters used to generate the synthetic spectra are those listed in Table~\ref{Tab:parameters}.
}
\end{figure*}

The wings of the Balmer lines were fitted, as well as those of strong lines like \ion{Ca}{I}~$\lambda$ 422.67~nm (Fig.~\ref{Fig:Ca}). 
Although the gravity and s-process overall abundances are entangled (s-process elements 
often involve lines from the singly ionized species, which are generally strengthened in low-density atmospheres), a solution with [s/Fe]~=~+2.0~dex\footnote
{This means that all heavy elements with the s-process contributing more than 50\% of their solar-system abundance were assigned an overabundance
of 2.0~dex in the spectral-synthesis computation.}  (or even more, except for the first s-process-peak element Y) and $\log g = 1$ represents a good overall match to the spectrum, as shown in Fig.~\ref{Fig:Ba} in a region containing many lines from singly-ionized s-process elements. We stress that the shape of the $^{12}$C$^{13}$C and  $^{13}$C$^{13}$C bandheads around 474.0 nm  are quite sensitive to gravity (see Fig.~\ref{Fig:C12C13}), and it is this criterion that led us to adopt $\log g = 1$.

\HE\ cannot be a low-metallicity carbon dwarf, since even at such a low metallicity, the wings of the more intense lines would be too strong. Fig.~\ref{Fig:Ca} reveals the impact of gravity on the   \ion{Ca}{I} $\lambda$~422.67~nm line. At higher gravities, the wings of the line become stronger, and metallicity has to be decreased (to [Fe/H]$ = -3.5$), but then other (metallic) lines start becoming really weak, and make this solution unacceptable.

\begin{figure}
\includegraphics[width=9cm]{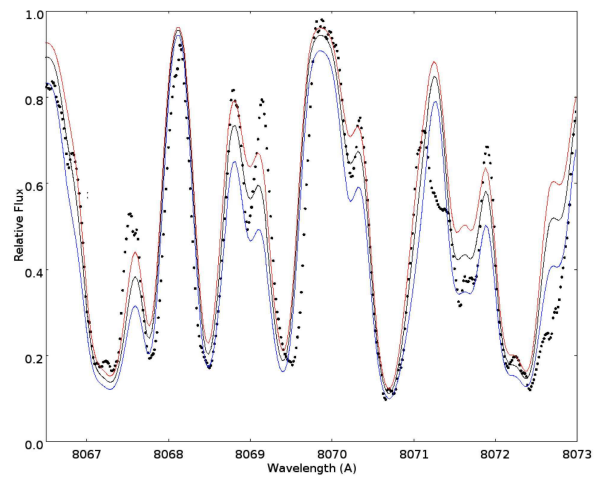}
\caption{\label{Fig:CN}
Intensities of lines from the CN band around $\lambda 8070$~\AA, showing the impact of the N abundance [with $\log\epsilon$(C) fixed at 8.2]:  $\log\epsilon$(N) = 7.7
(red curve), $\log\epsilon$(N) = 7.9
(black curve), $\log\epsilon$(N) = 8.2
(blue curve). Except otherwise stated, the atmospheric parameters used to generate the synthetic spectra are those listed in Table~\ref{Tab:parameters}.
}
\end{figure}

Especially noteworthy is the fact that \HE\ combines features from both J stars (a low $^{12}$C/$^{13}$C ratio, as inferred from the  $^{12}$C$^{13}$C and  $^{13}$C$^{13}$C bandheads around 474.0 nm; see Fig.~\ref{Fig:C12C13}) and CH stars (large s-process overabundances, and strong CH bands in the region 430 - 440~nm, caused by a large C overabundance and a very low metallicity [Fe/H] = -2.4). The high N abundance, as derived from the CN band longward of 570~nm or 800~nm (Fig.~\ref{Fig:CN}), is also remarkable, and is consistent with the low $^{12}$C/$^{13}$C ratio ($\sim 4$), since both features are signatures of CN-processing, as it may occur  along the red giant branch, with extra-mixing bringing these products to the stellar surface \citep{2014ApJ...797...21P}. The  N overabundance   ([N/H] = 0.07, [C+N/H] = 0.85, and [C/N] = 0.16) is in fact typical of CEMP-rs stars  (we will show in Sect.~\ref{Sect:sprocess} that \HE\ is indeed a CEMP-rs star), as may be seen from  Fig.~17 of \citet{2010A&A...509A..93M}. With the adopted atmospheric parameters, there seems to be no way of avoiding a strong N overabundance if the strengths of the CN lines are to be reproduced. The carbon abundance of \HE\   ([C/H] = -0.23) seems reasonable as well among CEMP-rs stars \citep[Fig.~14 from][]{2010A&A...509A..93M}.

In conclusion, \HE\ must thus be considered as another member of  the CEMP class, albeit one of the coolest and most luminous member of the class, 
but not as extreme as CS~30322-023, the latter having atmospheric parameters \Teff\ = 4100~K, $\log g \le   -0.3$,  and [Fe/H]$= -3$ \citep{2006A&A...455.1059M}. The atmospheric parameters of \HE\ are used in Sect.~\ref{Sect:sprocess} to derive abundances for s-process elements. The  evolutionary status of \HE\ is discussed next  (Sect.~\ref{Sect:evol}).

\begin{table*}
\caption{\label{Tab:parameters}
Adopted atmospheric parameters of \HE\ ($\xi_t$ is the microturbulence).}
\begin{tabular}{ccccccccccccc}
\hline
\Teff & $\log g$ & [Fe/H] & \multicolumn{3}{c}{$\log\epsilon$} &&  \multicolumn{3}{c}{[X/H]} & $^{12}$C/$^{13}$C & [s/Fe] & $\xi_t$\\
\cline{4-6}\cline{8-10}
(K) & & & C & N  & O&&C & N  & O & & & (\kms)\\
\hline
$4250\pm100$ & $1\pm1$ &  -2.4 & $8.2\pm0.1$ & $7.9\pm0.1$ &   7.1$^a$ && $-0.23$& $2.1$ & $-1.6^a$ & $4\pm1$ & $2.0\pm0.2$ & 2\\
\hline
\end{tabular}

$^a$  oxygen abundance imposed from [O/Fe] = 0.8 \citep[see Fig.~18 of][]{2010A&A...509A..93M}.
\end{table*} 

\section{Heavy-element abundances}
\label{Sect:sprocess}

A detailed analysis of  abundances for some key elements  (some typical of the s-process, others typical of the r-process) has been performed using the parameters from Table~\ref{Tab:parameters}, using the Keck/Hires spectrum obtained on October 7, 2008 and reduced using the standard "Makee" pipeline. The line list is given in Table~\ref{Tab:linelist}, avoiding spectral regions with the unaccounted SiC$_2$ Merrill-Sanford bands.  Oscillator strengths are from the VALD database \citep{2011BaltA..20..503K}. The corresponding abundances are listed in Table~\ref{Tab:abundances}.
The carbon-star model used has [Fe/H]= $-2$, \Teff \ = 4250~K, $\log g = 1.00$,  [$\alpha$/Fe] = +0.4, $\xi_t = 2.0$~\kms, $\log\epsilon$(C) = 8.2, $\log\epsilon$(N) = 7.4, $\log\epsilon$(O) = 7.1, [s/Fe] = +2.0~dex. The synthetic spectrum  was convolved with a macroturbulence velocity of 8.0~\kms.
 An analysis of the sensitivity of the Fe, Zr and Eu abundances to the atmospheric parameters is presented in 
Table~\ref{Tab:errors}.  The global uncertainty, adding up all individual uncertainties, is on the order of $\pm0.3$~dex for the heavy elements Zr and Eu.
This is somewhat larger than usual, but is nevertheless of the same order as the uncertainties found by \citet{2009A&A...508..909Z} 
in their analysis of R-type carbon stars, thus revealing the difficulties inherent in the analysis of carbon stars.

\begin{table}
\caption[]{\label{Tab:abundances}
Abundances (in the scale $\log\epsilon(\rm{H}) = 12$) of Fe and heavy elements, the line-to-line standard deviation is between parenthesis.  $N$ is the number of lines used in the computation 
of  the average abundance. 
Abundances have been normalized by the solar photospheric values from \citet{2009ARA&A..47..481A}. 
}
\begin{tabular}{lrrrrrrrrr}
\hline
X & \multicolumn{1}{c}{$\log\epsilon$(X)} & [X/H] & [X/Fe] & $N$\\
\hline
Fe & 5.02 (0.07) & -2.4 & - & 6 \\
Y II & 0.33 (0.03) &  -1.9  & 0.5 & 2\\
Zr II & 1.73 (0.04) & -0.9 & 1.6 & 4 \\
La II & 1.12 (0.13) & 0.0 &  2.4 & 5\\
Ce II & 1.22 (0.06) & -0.5 & 2.0 & 3\\
Nd II & 1.20 (0.06) & -0.3 & 2.2 & 7\\
Sm II & 0.43 (0.03) & -0.6 & 1.9 & 2\\
Eu II & 0.35 (0.05) & -0.2 & 2.3 & 2\\
Dy II & -0.10 (0.20) & -1.2 & 1.2 & 2\\
Er II & 0.25 (0.05) & -0.7 & 1.8 & 2\\
\hline
\end{tabular}
\end{table}

The correct Ce abundance was used when deriving the Li abundance, since the \ion{Ce}{II} line at 
$\lambda$~6708.099 blends the \ion{Li}{I} resonance doublet at $\lambda$~6707.706 and 6707.91 
\citep{2002A&A...395L..35R}.
Only an upper limit could be set on the Li abundance: $\log\epsilon$(Li)$\; < 1.0$. This star 
is thus not Li-rich.

As can be seen from Table~\ref{Tab:abundances},  the heavy-element abundance pattern  in \HE\ 
looks at first sight typical of a CEMP-s star, with moderate overabundances for elements belonging to the first s-process peak (Y, Zr), and stronger overabundances for the second peak (La, Ce, Nd). However, the analysis of the r-process elements Sm, Eu, Dy and Er reveals that they are enriched as well, and therefore \HE\ is likely a CEMP-rs star. The defining criterion of a CEMP-rs star is based on the [Ba/Eu] ratio being in the range 0.0 -- +0.5~dex, 
whereas [Ba/Eu]~$ > 0.5$~dex for CEMP-s stars \citep[][see also Figs.~1 and 2 of Masseron et al. 2010]{2005ARA&A..43..531B}. Unfortunately, 
the Ba abundance could not be derived safely, because the 493.407, 585.37 and 614.14~nm lines were either blended or with strong and badly fitted wings. Nevertheless, a lower limit on the Ba abundance could be set at $\log\epsilon$(Ba)~$ > 1.6$~dex (using  the 493.407 and 614.14~nm lines), corresponding to 
[Ba/H]~$ > -0.6$ and  [Ba/Fe]~$ > 1.9$. Combined with [Eu/Fe] = 2.3~dex (Table~\ref{Tab:abundances}), these values confirm the CEMP-rs nature of \HE\ \citep[see also Fig.~1 of][]{2010A&A...509A..93M}.
In that respect, it is noteworthy that the [La/Ce] ratio (0.5~dex) is much larger than the typical (negative) values predicted by the computations of  \citet{2005...GS} 
from the operation of the $^{13}$C($\alpha$, n)$^{16}$O neutron source in low-mass AGB stars. The
 larger [La/Ce] values observed in \HE\ as in all the other CEMP-rs stars
are thus a strong indication that the $^{13}$C($\alpha$, n)$^{16}$O neutron source
does not operate in those stars, as suggested by \citet{2010A&A...509A..93M}.
It is also meaningful that the [La/Ce]
values observed in CEMP-rs stars are compatible with the values
0.2–0.4 dex predicted from the operation of the $^{22}$Ne($\alpha$, n)$^{25}$Mg
neutron source in warm pulses, and after dilution in the AGB
envelope  \citep{2005...GS}.

Unfortunately, the abundance  of Pb, falling in the third s-process peak  and generally  much enhanced in CEMP-rs stars, could not be easily derived, as the Pb blend is not cleanly reproduced. 

Only an upper limit could be set on the technetium abundance ($\log\epsilon$(Tc)$ < 0.0$), which hints at the absence of Tc, thus confirming the extrinsic nature
of \HE.

\section{The evolutionary status of HE 0017+0055}
\label{Sect:evol}

\begin{figure}
\centering \includegraphics[width=8cm]{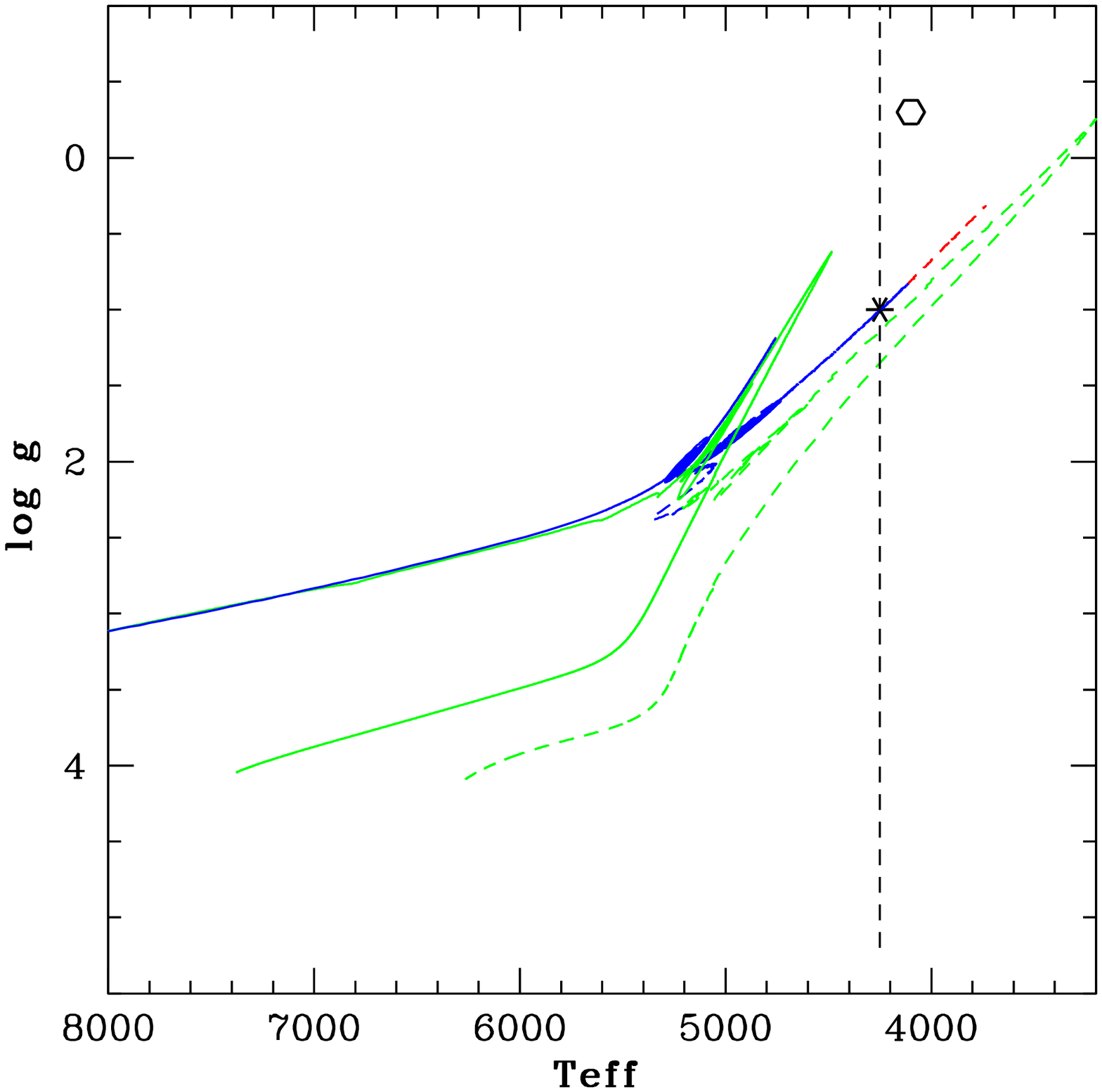}
    \caption{
Evolutionary track in the (\Teff, $\log g$) diagram for a 0.9~\Msun\ star of metallicity [Fe/H]~=~ -2.4, computed with the STAREVOL code, with an atmosphere 
being either oxygen-rich (solid track) or carbon-rich (dashed track, corresponding to $\log\epsilon(C) = 8.2$ or [C/H] $= -0.23$). The oxygen-rich track is limited 
to the giant and horizontal branches (green and blue parts of the track, respectively), for the sake of clarity, whereas the carbon-rich track includes as well the 
asymptotic giant branch (red part of the track).
The vertical dashed line corresponds to \Teff\ ~=~4250~K. The location of \HE\ as inferred from the parameters of Table~\protect\ref{Tab:parameters} is indicated by the star symbol, whereas that of CS~30322-023 is indicated by the open circle.
  }
  \label{Fig:HR}
\end{figure}

Being of very low metallicity, the star must have formed in the early stage of the Galaxy: 
according to Fig.~2 of \citet{2002ARAA...40..487F}, stars with metallicities [Fe/H]$ = -2.4$ like \HE\ 
formed about 13~Gyr ago. Since \HE\ is located on the giant branch, it must have spent 
about 12~Gyr on the main sequence (and the remainder in post-main-sequence phases) 
and must therefore have had an initial mass of about 0.9~\Msun.
The gravity and effective temperatures of \HE\ mentioned in
Sect.~\ref{Sect:parameters} may be combined, along with the above mass estimate (0.9~\Msun), to 
derive the luminosity:
\begin{equation}
\log L/L_{\odot} = \log M/M_\odot +  4 \log T_{\rm eff}/T_{\rm eff,\odot} - \log g/g_\odot = 2.86,  
\end{equation}
corresponding to $L = 724$~L$_{\odot}$, in agreement with the prediction of the STAREVOL C-rich model described below (and Table~\ref{Tab:space}). 
According to the compilation of CEMP-star properties by 
\citet{2010A&A...509A..93M}, this value for the luminosity makes \HE\ one the brightest CEMP stars 
(their Fig.~12), just below CS~30322-023 \citep[Fig.~\ref{Fig:HR} and ][]{2006A&A...455.1059M}. The radius corresponding to \HE\ gravity and adopted mass is 50~\Rsun. Only CS~30322-023, with $\log g = -0.3\pm0.3$ and \Teff\ = 4100 - 4350~K \citep{2006A&A...455.1059M},
has a higher luminosity among CEMP stars. 
Although CS~30322-023 most likely lies above the RGB tip (Fig.~\ref{Fig:HR}), and 
might therefore belong to the rare class of low-metallicity AGB stars, this is not required for \HE.

Fig.~\ref{Fig:HR} presents evolutionary tracks of a 0.9~\Msun\ star with a metallicity [Fe/H]$ = -2.4$ like \HE\  computed with the STAREVOL code \citep{2008A&A...489..395S}. The exact location of the giant branch in the Hertzsprung-Russell diagram is sensitive to the opacity in the atmosphere, and hence to the carbon- or oxygen-rich nature of the star \citep{2002A&A...387..507M}. For a star strongly enriched in carbon as is \HE , the evolutionary  track is shifted 
towards lower temperatures, thus counteracting to some extent the blueward shift due to the low metallicity. This effect appears crucial to derive atmospheric parameters consistent with the ones derived from the spectral analysis in Sect.~\ref{Sect:re-analysis}, and especially to avoid the high luminosities that would yield large space motions.

The distance of \HE\ may then be estimated, adopting the same bolometric correction of 0.3 as that estimated by
\citet{2006A&A...455.1059M} for \CS, a distance of 4.8~kpc is obtained (Table~\ref{Tab:space}).  At such a distance, the Tycho-2 proper motions (see Sect.~\ref{Sect:kinematics}) translate into 
large space velocities as listed in Table~\ref{Tab:space}\footnote{A similar analysis is unfortunately not 
possible for \CS\ for which no proper motion is available.}. As discussed in Sect.~\ref{Sect:kinematics}, these space velocities are consistent with \HE\ being a normal member of the Galactic halo.

 An accurate trigonometric parallax  for \HE\ should become available 
from Gaia within a few years time and should provide a conclusive answer to 
the discussion on the distance and evolutionary stage of this star. 

\begin{table*}
\caption[]{\label{Tab:errors}
 Sensitivity of the derived abundances (in dex)  on the atmospheric parameters.
For non-symmetrical abundance variations, the largest value is listed.
The final uncertainty on the abundance is computed by quadratically adding all the uncertainties,  including the
standard deviation from the line-to-line scatter. 
}
\begin{tabular}{lllllll}
\hline
Element & $T_{\rm eff}$                & $\log g$       & [Fe/H]                 & $\xi$                               & macro.                          & Total \\
              & $\pm 100$~K & $\pm 0.5$   & $\pm 0.25$~dex & $\pm 1$ km s$^{-1}$   & $\pm 1$~km s$^{-1}$ &     (dex)       \\
\hline
Fe        & $\pm 0.09$ & $\pm 0.07$   &   --                   &  $\pm 0.12$    & $\pm 0.03 $    & $\pm 0.18$  \\
Zr         & $\pm 0.20$ & $\pm 0.11$   & $\pm 0.07$   &  $\pm 0.08 $   & $\pm 0.11  $   & $\pm 0.29$  \\
Eu        & $\pm 0.20$ & $\pm 0.15 $  & $\pm 0.20$   &  $\pm 0.10 $   & $\pm 0.07  $   & $\pm 0.34$  \\
\hline
\end{tabular}
\end{table*}

\section{Conclusions}

The radial-velocity monitoring of \HE\ revealed short-term variations superimposed on a long-term trend. The short-term variations are characterised by a period of 383~d and a very small mass function of $(6.1\pm1.1)\times10^{-6}$~\Msun. There are three possible explanations for these low-amplitude variations: (i) 
orbital variations caused by a WD companion in a very inclined orbit ($i \sim 2.3^\circ$), (ii) orbital variations caused  by  a brown dwarf in a moderately inclined orbit ($12.4^\circ \le i \le 90^\circ$), (iii) velocity jitter reported by \citet{2003AJ....125..293C} as typical in low-metallicity giants close to the RGB tip. \citet{2015A&A...XXX..NNNJ} report three more cases among CH/CEMP-s systems 
of small-amplitude velocity variations with periods close to 1~yr, very similar thus to the one observed in \HE. A definite conclusion as to which possibility is the valid one must await the results of an accurate photometric monitoring, since the variations are expected to be minute (a few hundredths of a magnitude).

An abundance study has shown that \HE\ is  not only enriched in s-process elements but also in r-process elements (at levels on the order of 2~dex for most of them), and may thus be flagged as a CEMP-rs star. A mass-transfer scenario is generally invoked to account for the abundance peculiarities of CEMP-rs stars. In this scenario, the CEMP-rs star has been polluted by  heavy-element-rich matter coming from a companion  formerly on the thermally-pulsing AGB star \citep[most probably a massive one;][]{2010A&A...509A..93M}, now a WD. In case (i) above, the WD is in the 
inner orbit, and the outer orbit hosts an innocent bystander star, whereas in cases (ii) and (iii), the WD is in the long-period orbit.
Although the outer orbit must have a period of several decades, this is still compatible with the mass-transfer scenario, since other s-process-polluted systems 
with similarly long orbital periods are known (like HD 26; Jorissen et al., 2015).  

\begin{acknowledgement}
This research has been funded by the Belgian Science Policy Office under contract BR/143/A2/STARLAB. 
SvE and LS are FNRS research associates. The work of T.T.H. was supported by Sonderforschungsbereich SFB 
881 ``The Milky Way System'' (subproject A4) of the German Research 
Foundation (DFG). J.A. and B.N. gratefully acknowledge financial support 
from the Danish Natural Science Research Council and the Carlsberg 
Foundation. 
Based on observations obtained with the
HERMES spectrograph, supported by the Fund for Scientific Research of Flanders (FWO),
the Research Council of K.U.Leuven, the Fonds National de la Recherche Scientifique
(F.R.S.-FNRS), Belgium, the Royal Observatory of Belgium, the Observatoire de Gen\`eve,
Switzerland and the Th\"uringer Landessternwarte Tautenburg, Germany. 
\end{acknowledgement}

\appendix
\section{Line list for the abundance analysis}

\begin{table}
\caption[]{\label{Tab:linelist}
Lines used in the metallicity and s-process abundance analysis, and the corresponding abundances (in the scale $\log\epsilon(\rm{H}) = 12$). The bold values at the end of each element's list  
correspond to the average abundance and its corresponding standard dispersion.
A vertical bar to the left of the wavelength values groups hyperfine structure and isotopic shifts for a given line.
}
\begin{tabular}{llrll}
\hline
 & \multicolumn{1}{c}{$\lambda$} & $\chi_{\rm exc}$& $\log gf$ & $\log\epsilon$  \\
 &  \multicolumn{1}{c}{(\AA)} & (eV) \\
\hline

\hline
Fe I & 4466.551 & 2.831 & -0.600 & 4.9 \\
& 4903.310 & 2.882 & -0.926 & 5.0 \\
& 5195.472 & 4.220 & -0.086 & 5.1 \\
& 6219.281 & 2.198 & -2.434 & 5.0 \\
& 6839.830 & 2.559 & -3.350 & 5.0 \\
& 7760.897 & 5.486 & -3.838 & 5.1 \\
& & & &  {\bf 5.02 (0.07)} \\
\medskip\\

Y II &  5200.406 & 0.992 & -0.570 & 0.35 \\
     &  5205.724 & 1.033 & -0.193 & 0.3 \\
     & & & &  {\bf 0.325 (0.025)}\\
\medskip\\

Zr II & 4457.413 & 1.184 & -1.220 & 1.7 \\
      & 4816.500 & 1.011 & -2.000 & 1.8 \\   
      & 4831.327 & 1.208 & -1.720 & 1.7 \\    
      & 4962.310 & 0.972 & -2.000 & 1.7 \\    
      & & & &  {\bf 1.73 (0.04) } \\
%
\medskip\\
La II & 4558.457 & 0.321 & -0.970 & 0.9\\
      & 4808.996 & 0.235 & -1.40  & 1.3\\
      & 4824.052 & 0.651 & -0.87  & 1.1 \\
      & \multicolumn{1}{|l}{4920.965} & 0.126 & -2.261 & \\
      & \multicolumn{1}{|l}{4920.965} & 0.126 & -2.407 & \\
      & \multicolumn{1}{|l}{4920.966} & 0.126 & -2.065 & \\
      & \multicolumn{1}{|l}{4920.966} & 0.126 & -2.078 & \\
      & \multicolumn{1}{|l}{4920.966} & 0.126 & -2.738 & \\
      & \multicolumn{1}{|l}{4920.968} & 0.126 & -1.831 & \\
      & \multicolumn{1}{|l}{4920.968} & 0.126 & -1.956 & \\
      & \multicolumn{1}{|l}{4920.968} & 0.126 & -2.629 & \\
      & \multicolumn{1}{|l}{4920.971} & 0.126 & -1.646 & \\
      & \multicolumn{1}{|l}{4920.971} & 0.126 & -1.895 & \\
      & \multicolumn{1}{|l}{4920.971} & 0.126 & -2.650 & \\
      & \multicolumn{1}{|l}{4920.975} & 0.126 & -1.490 & \\
      & \multicolumn{1}{|l}{4920.975} & 0.126 & -1.891 & \\
      & \multicolumn{1}{|l}{4920.975} & 0.126 & -2.760 & \\
      & \multicolumn{1}{|l}{4920.979} & 0.126 & -1.354 & \\
      & \multicolumn{1}{|l}{4920.979} & 0.126 & -1.957 & \\
      & \multicolumn{1}{|l}{4920.979} & 0.126 & -2.972 & \\
      & \multicolumn{1}{|l}{4920.985} & 0.126 & -1.233 & \\
      & \multicolumn{1}{|l}{4920.985} & 0.126 & -2.162 & \\
      & \multicolumn{1}{|l}{4920.985} & 0.126 & -3.375 & 1.2
\medskip\\
\hline
\end{tabular}
\end{table}

\addtocounter{table}{-1}
\begin{table}
\caption{Continued}
\begin{tabular}{llrll}
\hline
 & \multicolumn{1}{c}{$\lambda$} & $\chi_{\rm exc}$& $\log gf$ & $\log\epsilon$  \\
 &  \multicolumn{1}{c}{(\AA)} & (eV) \\
\hline
\medskip\\
La II   & \multicolumn{1}{|l}{4921.774} & 0.244 & -1.139 & \\
      & \multicolumn{1}{|l}{4921.774} & 0.244 & -2.220 & \\
      & \multicolumn{1}{|l}{4921.774} & 0.244 & -3.601 & \\
      & \multicolumn{1}{|l}{4921.775} & 0.244 & -1.233 & \\
      & \multicolumn{1}{|l}{4921.775} & 0.244 & -2.005 & \\
      & \multicolumn{1}{|l}{4921.775} & 0.244 & -3.207 & \\
      & \multicolumn{1}{|l}{4921.776} & 0.244 & -1.334 & \\
      & \multicolumn{1}{|l}{4921.776} & 0.244 & -1.445 & \\
      & \multicolumn{1}{|l}{4921.776} & 0.244 & -1.915 & \\
      & \multicolumn{1}{|l}{4921.776} & 0.244 & -1.927 & \\
      & \multicolumn{1}{|l}{4921.776} & 0.244 & -2.923 & \\
      & \multicolumn{1}{|l}{4921.776} & 0.244 & -3.010 & \\
      & \multicolumn{1}{|l}{4921.777} & 0.244 & -1.566 & \\
      & \multicolumn{1}{|l}{4921.777} & 0.244 & -1.955 & \\
      & \multicolumn{1}{|l}{4921.777} & 0.244 & -2.939 & \\
      & \multicolumn{1}{|l}{4921.778} & 0.244 & -1.700 & \\
      & \multicolumn{1}{|l}{4921.778} & 0.244 & -1.848 & \\
      & \multicolumn{1}{|l}{4921.778} & 0.244 & -2.006 & \\
      & \multicolumn{1}{|l}{4921.778} & 0.244 & -2.053 & \\
      & \multicolumn{1}{|l}{4921.778} & 0.244 & -2.258 & \\
      & \multicolumn{1}{|l}{4921.778} & 0.244 & -3.123 &  1.1\\
& & & &  {\bf 1.12 (0.13)}\\
\medskip\\
Ce II & 4835.674 & 0.957  & -0.870 & 1.2 \\
      & 5187.458 & 1.212  & 0.150 & 1.15 \\ 
      & 5191.633 & 0.869  &-0.560 & 1.3\\
& & & &  {\bf 1.22 (0.06)}\\
\medskip\\
Nd II & 4594.447 & 0.205 & -1.360 & 1.2 \\ 
      & 4820.339 & 0.205 & -0.920 & 1.28\\
      & 4825.478 & 0.182 & -0.420 & 1.2 \\
      & 4859.026 & 0.321 & -0.440 & 1.1 \\
      & 5228.420 & 0.380 & -1.280 & 1.3 \\ 
      & 5249.576 & 0.976 &  0.200 & 1.2 \\ 
      & 5276.869 & 0.859 & -0.440 & 1.15 \\
      & & & &  {\bf 1.20 (0.06)}\\
\medskip\\
Sm II &   4777.840 & 0.040 & -1.420 & 0.45 \\
      &   4815.800 & 0.185 & -0.820 & 0.4 \\
      & & & &  {\bf 0.43 (0.03) }\\
\hline
\end{tabular}
\end{table}

\addtocounter{table}{-1}
\begin{table}
\caption{Continued}
\begin{tabular}{llrll}
\hline
 & \multicolumn{1}{c}{$\lambda$} & $\chi_{\rm exc}$ & $\log gf$ & $\log\epsilon$  \\
 &  \multicolumn{1}{c}{(\AA)} & (eV) \\
\hline
\medskip\\
Eu II &  \multicolumn{1}{|l}{4522.486} & 0.207  & -2.159 &     \\ 
      &  \multicolumn{1}{|l}{4522.498} & 0.207  & -1.266 &     \\
      &  \multicolumn{1}{|l}{4522.538} & 0.207  & -1.984 &     \\
      &  \multicolumn{1}{|l}{4522.549} & 0.207  & -1.474 &     \\
      &  \multicolumn{1}{|l}{4522.561} & 0.207  & -2.159 &     \\
      &  \multicolumn{1}{|l}{4522.581} & 0.207  & -1.962 &     \\
      &  \multicolumn{1}{|l}{4522.590} & 0.207  & -1.711 &     \\
      &  \multicolumn{1}{|l}{4522.600} & 0.207  & -1.984 &     \\
      &  \multicolumn{1}{|l}{4522.615} & 0.207  & -2.038 &     \\
      &  \multicolumn{1}{|l}{4522.621} & 0.207  & -1.980 &     \\
      &  \multicolumn{1}{|l}{4522.630} & 0.207  & -1.962 &     \\
      &  \multicolumn{1}{|l}{4522.640} & 0.207  & -2.247 &     \\
      &  \multicolumn{1}{|l}{4522.644} & 0.207  & -2.256 &     \\
      &  \multicolumn{1}{|l}{4522.650} & 0.207  & -2.038 &     \\
      &  \multicolumn{1}{|l}{4522.657} & 0.207  & -2.344 &     \\
      &  \multicolumn{1}{|l}{4522.661} & 0.207  & -2.247 & 0.3   
\medskip\\
      &  \multicolumn{1}{|l}{6437.637} & 1.320  & -1.428 &     \\
      &  \multicolumn{1}{|l}{6437.639} & 1.320  & -2.206 &     \\
      &  \multicolumn{1}{|l}{6437.635} & 1.320  & -2.206 &     \\
      &  \multicolumn{1}{|l}{6437.637} & 1.320  & -1.377 &     \\
      &  \multicolumn{1}{|l}{6437.637} & 1.320  & -2.010 &     \\
      &  \multicolumn{1}{|l}{6437.635} & 1.320  & -2.010 &     \\
      &  \multicolumn{1}{|l}{6437.635} & 1.320  & -1.287 &     \\
      &  \multicolumn{1}{|l}{6437.633} & 1.320  & -1.956 &     \\
      &  \multicolumn{1}{|l}{6437.633} & 1.320  & -1.956 &     \\
      &  \multicolumn{1}{|l}{6437.630} & 1.320  & -1.181 &     \\
      &  \multicolumn{1}{|l}{6437.623} & 1.320  & -1.998 &     \\
      &  \multicolumn{1}{|l}{6437.627} & 1.320  & -1.998 &     \\
      &  \multicolumn{1}{|l}{6437.620} & 1.320  & -1.070 &     \\
      &  \multicolumn{1}{|l}{6437.606} & 1.320  & -2.191 &     \\
      &  \multicolumn{1}{|l}{6437.617} & 1.320  & -2.191 &     \\
      &  \multicolumn{1}{|l}{6437.603} & 1.320  & -0.960 & 0.4    \\
      & & & &  {\bf 0.35 (0.05)}\\
\medskip\\

Dy II &  3984.688 & 1.752  & -0.328 & +0.1   \\
      &  3996.689 & 0.590  & -0.260 & -0.3   \\
      & & & & {\bf  -0.10 (0.20) }\\
\medskip\\

Er II &  3980.144 & 0.886  & -1.004 & 0.2    \\
      &  4008.177 & 0.055  & -1.547 & 0.3    \\
            & & & & {\bf  0.25 (0.05) }\\
\hline
\end{tabular}
\end{table}

\end{document}